# Intracluster Moves for Constrained Discrete-Space MCMC

**Firas Hamze**
D-Wave Systems Inc.
100-4401 Still Creek Drive
Burnaby, B.C., V5C 6G9, Canada

**Nando de Freitas**
Department of Computer Science
University of British Columbia
Vancouver, B.C., Canada

## Abstract

This paper addresses the problem of sampling from binary distributions with constraints. In particular, it proposes an MCMC method to draw samples from a distribution of the set of all states at a specified distance from some reference state. For example, when the reference state is the vector of zeros, the algorithm can draw samples from a binary distribution with a constraint on the number of active variables, say the number of 1's. We motivate the need for this algorithm with examples from statistical physics and probabilistic inference. Unlike previous algorithms proposed to sample from binary distributions with these constraints, the new algorithm allows for large moves in state space and tends to propose them such that they are energetically favourable. The algorithm is demonstrated on three Boltzmann machines of varying difficulty: A ferromagnetic Ising model (with positive potentials), a restricted Boltzmann machine with learned Gabor-like filters as potentials, and a challenging three-dimensional spin-glass (with positive and negative potentials).

## 1 INTRODUCTION

Sampling from binary distributions with constraints on the number of active variables is a problem of interest in statistical physics and probabilistic inference. In physics, the problem arises when adopting binary probabilistic graphical models, known as *conserved-order-parameter Ising models*, to study the properties of lattice gases (Newman & Barkema 1999, Chapter 5). Here, fixing the number of active variables has the physical meaning of holding the magnetization of the system constant.

In probabilistic inference, constraints on the number of active binary hidden variables in a Boltzmann machine are often imposed to either control the complexity of the coding scheme, regularize the problem or simulate the effect of lateral inhibition in hierarchical models of perception (Kappen 1995). In the model adopted by Kappen the latent variables are independent of each other given the observed variables and, hence, the constraint on the latent variables introduces interaction among them. This type of restricted Boltzmann machine, albeit without this precise interaction term, has become a popular building block in deep architectures (Hinton & Salakhutdinov 2006).

There is also a remarkably large number of other statistical inference problems, where one can apply Rao-Blackwellization (Hamze & de Freitas 2004) to integrate out all continuous variables and end up with a discrete distribution. Examples include topic modeling and Dirichlet processes (Blei, Ng & Jordan 2003), Bayesian variable selection (Tham, Doucet & Kotagiri 2002), mixture models (Liu, Zhang, Palumbo & Lawrence 2003) and multiple instance learning (Kück & de Freitas 2005). In these domains, it might sometimes be desirable to impose constraints on the number of active topics, clusters, variables or models. If such an avenue is pursued, then the MCMC method proposed in this paper would provide a reasonable inference engine for these constrained discrete sampling problems.

This paper presents a specialized algorithm for equilibrium Monte Carlo sampling of binary-valued systems with constraints. In particular, the proposed MCMC method draws samples from a distribution defined on the set of all states *at a specified distance* from a reference state. The method allows for large moves in the state space. That is, many bits are flipped in a single MCMC step.

Despite a great deal of interest on this problem in statistical physics, by far the most popular existing algorithms are based on single pair-wise bit ex-change moves. Examples include the Kawasaki algorithm and the improved Metropolis bit-swap algorithm (Kawasaki 1966, Newman & Barkema 1999) that we discuss in the next section. As stated in (Newman & Barkema 1999, Section 5.1), neither single-variable-flip algorithms nor cluster algorithms, such as Swendsen-Wang, can be easily applied to conserved-order-parameter Ising models because it is hard to ensure that the magnetization constraint is not violated. The MCMC moves proposed here will allow for many, as opposed to a single pair of, variables to change at each iteration without violating the magnetization constraint.

We should also point out that sequential Monte Carlo (SMC) methods, such as hot coupling and annealed importance sampling, have been successfully used to sample from Boltzmann machines without constraints (Hamze & de Freitas 2005, Salakhutdinov & Murray 2008). Since such samplers often use an MCMC kernel as proposal distribution, the MCMC sampler proposed in this paper would be useful to extend the domain of application of these SMC methods to Boltzmann machines with constraints of the type considered here.

# 2 PRELIMINARIES

Consider a binary-valued system defined on the state space $\mathcal{S} \triangleq \{0,1\}^M$, *i.e.* consisting of $M$ variables each of which can be 0 or 1. The probability of a state $\mathbf{x}$ is given by the Boltzmann distribution:

$$\pi(\mathbf{x}) = \frac{1}{Z(\beta)} e^{-\beta E(\mathbf{x})}$$

where $\beta$ is an *inverse temperature.* An instance of such a system is the ubiquitous *Ising model* of statistical physics, also called a *Boltzmann machine* by the machine learning community (Ackley, Hinton & Sejnowski 1985).

Consider a particular state $\mathbf{c}$, which for now can be considered arbitrary. We are interested in drawing samples from the set of all states at Hamming Distance $n$ from $\mathbf{c}$, which we call $\mathcal{S}_n(\mathbf{c})$. For example if $M = 3$ and $\mathbf{c} = [1,1,1]$, then $\mathcal{S}_0(\mathbf{c}) = \{[1,1,1]\}$, $\mathcal{S}_1(\mathbf{c}) = \{[0,1,1],[1,0,1],[1,1,0]\}$, etc. Clearly, $|\mathcal{S}_n(\mathbf{c})| = \binom{M}{n}$. The partition function on $\mathcal{S}_n(\mathbf{c})$ is given by:

$$Z_n(\mathbf{c}) = \sum_{\mathbf{x} \in \mathcal{S}_n(\mathbf{c})} e^{-\beta E(\mathbf{x})}$$

$\mathcal{S}$ and $Z_n$ are understood to always depend on $\mathbf{c}$, so we can drop the explicit dependence in our notation. The *restricted* distribution we want to sample from is given by

$$\pi_n(\mathbf{x}) \triangleq \begin{cases} \frac{1}{Z_n} e^{-\beta E(\mathbf{x})} & \mathbf{x} \in \mathcal{S}_n \\ 0 & \text{otherwise} \end{cases} \quad (1)$$

In particular, when $\mathbf{c}$ is the vector of zeros, the distribution $\pi_n(\mathbf{x})$ has $n$ variables set to 1. This corresponds to the constant magnetization constraint mentioned in the introduction, but it is clear that the proposed sampling method applies to more general constraints. Nonetheless, in all subsequent examples in the algorithm description, we will assume for notational simplicity that $\mathbf{c}$ is the vector of zeros.

We define the sets of bits in state $\mathbf{x}$ that agree and disagree with those of $\mathbf{c}$: let $\mathcal{P}(\mathbf{x}) = \{i | x_i = c_i\}$ and $\mathcal{N}(\mathbf{x}) = \{i | x_i \neq c_i\}$. Thus, $\mathcal{P}(\mathbf{x}) \cup \mathcal{N}(\mathbf{x}) = \{1, \ldots M\}$ and $\mathcal{P}(\mathbf{x}) \cap \mathcal{N}(\mathbf{x}) = \emptyset$. Another useful definition is that of the *flip operator*, which simply inverts bit $i$ in a state, $F(\mathbf{x}, i) \triangleq (x_1, \ldots, \bar{x}_i, \ldots, x_M)$. We will rely heavily on this notation when defining the proposal moves.

In order to sample from the distribution (1), a possible Metropolis-type algorithm is to propose to flip, with uniform probability, a bit from the set and unset bits respectively, and accept the move (swap) with the Metropolis accept rule. For example, if $\mathbf{x} = [1,1,1,0,0,0,0]$, we may propose state $\mathbf{x}' = [1,1,0,0,1,0,0]$, and accept it with probability:

$$\alpha = \min(1, e^{-\beta(E(\mathbf{x}') - E(\mathbf{x}))})$$

This is a popular sampling scheme for these models (Newman & Barkema 1999, Hadjiagapiou, Malakis & Martinos 2006). While in principle correct, it can easily suffer from the usual troubles that beset MCMC algorithms, namely the issue of local minima. The set of allowable moves from $\mathbf{x}$ define its state space *neighbors*, and if all such neighbors have a higher energy than that of $\mathbf{x}$, then at low temperatures (large $\beta$) the sampler will remain at $\mathbf{x}$ for a long time. In the next section we detail a potential way around this issue.

# 3 INTRACLUSTER MOVES

We begin by describing our novel Monte Carlo move that allows for large changes of state *and* tends to propose them such that they are energetically favourable. Subsequently, we show that the resultant algorithm does in fact satisfy the theoretical requirements that ensure correct asymptotic sampling. We call the proposal the *intracluster move* (IM) as it generates states within the "cluster" $\mathcal{S}_n$. The name should not, however, generate confusion with "cluster-flipping" algorithms, such as that of Swendsen and Wang (Swendsen & Wang 1987, Gore & Jerrum 1997), which are unrelated.

Suppose that we have a state $\mathbf{x}_0$ on $\mathcal{S}_n$. It is possible to imagine taking special types of biased *self-avoiding walk* (SAW) of length $k$ *in the state space*, in other words sequences of states such that no state recurs in the sequence. The first type we consider, where states on sets $\{\mathcal{S}_{n-1} \dots \mathcal{S}_{n-k}\}$ are visited consecutively, is referred to as an *upward* SAW; the analogous case, where the walk is set to visit states on $\{\mathcal{S}_{n+1} \dots \mathcal{S}_{n+k}\}$, is called a *downward* SAW. In the ensuing description, we shall detail a joint move consisting of an upward SAW followed by a downward one; the case of a move with the two SAW types appearing in reverse order is straightforward once the principles are understood.

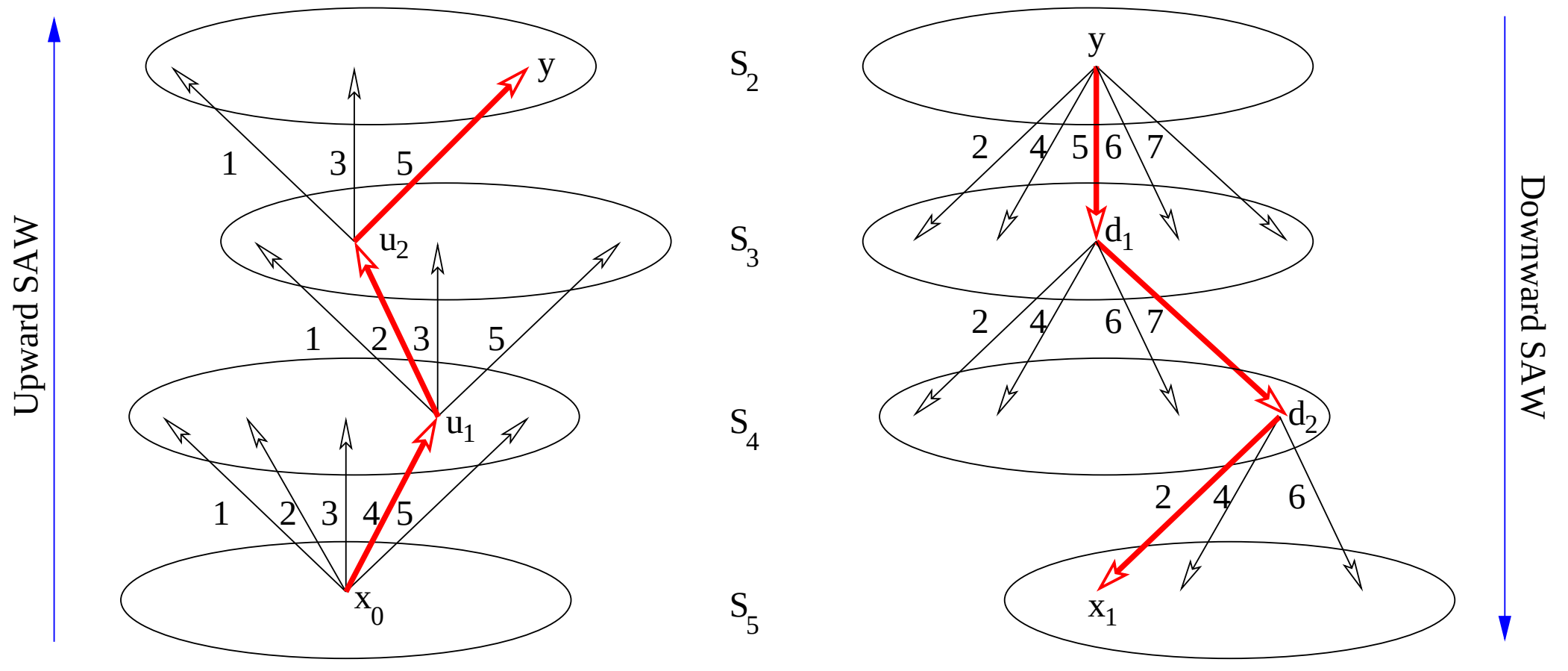

Figure 1: A visual illustration of the move process for up/down SAWs of length 3. The arrows represent the allowable moves from a state at that step; the red arrow shows the actual move taken in this example. From the system at state $\mathbf{x}_0 = [1,1,1,1,1,0,0]$ on $\mathcal{S}_5$, the *upward* SAW begins. Bit 4 of $\mathbf{x}_0$ is flipped to yield state $\mathbf{u}_1 = [1,1,1,0,1,0,0]$; the process is repeated until state $\mathbf{y} = [1,0,1,0,0,0,0]$ on $\mathcal{S}_2$ is reached. From there, the *downward* SAW considers and selects sequences of states in an analogous manner (in this case, $\mathbf{d}_1 = [1,0,1,0,1,0,0]$ is visited from $\mathbf{y}$, etc.) until the final state $\mathbf{x}_1 = [1,1,1,0,1,0,1]$ is reached. The sequence of states taken by the upward and downward SAWs are, respectively, $\boldsymbol{\sigma} = [4,2,5]$ and $\boldsymbol{\rho} = [5,7,2]$.

From $\mathbf{x}_0$, the $n$ single-flip neighbors on $\mathcal{S}_{n-1}$ are specified by the set $\mathcal{P}(\mathbf{x}_0)$. We may select a bit to flip, which we call $\sigma_1$, from this set in an energy-biased manner as follows:

$$f_{up}(\sigma_1 = i|\mathbf{x}_0) = \begin{cases} \frac{e^{-\gamma E(F(\mathbf{x}_0,i))}}{\sum_{j\in\mathcal{P}(\mathbf{x}_0)} e^{-\gamma E(F(\mathbf{x}_0,j))}} & i \in \mathcal{P}(\mathbf{x}_0) \\ 0 & \text{otherwise} \end{cases} \tag{2}$$

The larger the value of the simulation parameter $\gamma$, the more likely the proposal (2) is to sample the lower-energy neighbors of $\mathbf{x}_0$; conversely if it is zero, a neighbor on $\mathcal{S}_{n-1}$ is selected completely at random. In principle, the value of $\gamma$ will be seen to be arbitrary; indeed it can even be different at each step of the SAW. This will be discussed further in Section 4, where we set $\gamma$ to be close to the inverse temperature $\beta$.

From the resultant state on $\mathcal{S}_{n-1}$, the process can be repeated to yield a state on $\mathcal{S}_{n-2}$ and so on until some state $\mathbf{y}$ on set $\mathcal{S}_{n-k}$ is visited via the sequence of bit flips $\boldsymbol{\sigma} \triangleq (\sigma_1, \dots \sigma_k)$. At that point the upward SAW is said to be terminated, and the downward SAW may begin. In a completely analogous manner to the upward case, we may select a neighbor of $\mathbf{y}$ on $\mathcal{S}_{n-k+1}$ by flipping bit $\rho_1$ of $\mathcal{N}(\mathbf{y})$ according to:

$$f_{down}(\rho_1 = i|\mathbf{y}) = \begin{cases} \frac{e^{-\gamma E(F(\mathbf{y},i))}}{\sum_{j\in\mathcal{N}(\mathbf{y})} e^{-\gamma E(F(\mathbf{y},j))}} & i \in \mathcal{N}(\mathbf{y}) \\ 0 & \text{otherwise} \end{cases}$$

and so forth to yield the point $\mathbf{x}_1$ back on $\mathcal{S}_n$ via the flip sequence $\boldsymbol{\rho} \triangleq (\rho_1, \dots \rho_k)$. Again the value of $\gamma$ is tunable as it was in the upward case. The state $\mathbf{y}$ is said to be a *bridge point* on $\mathcal{S}_{n-k}$ of $\mathbf{x}_0$ and $\mathbf{x}_1$.

The sequence of *states* visited in the up and down processes are denoted, respectively, by $(\mathbf{u}_0, \dots, \mathbf{u}_k)$ and $(\mathbf{d}_0, \dots, \mathbf{d}_k)$, with $\mathbf{u}_0 = \mathbf{x}_0$, $\mathbf{u}_k = \mathbf{d}_0 = \mathbf{y}$, and $\mathbf{d}_k = \mathbf{x}_1$. The set of bits that can flip with nonzero probability are called the *allowable moves* at each step. A diagrammatic depiction of the up/down SAWs is shown in Figure 1. At this point it should be clear why the term "self-avoiding" aptly describes the up and down processes; by construction a state can never occur twice at any stage within each process, though of course in principle one that has been visited by the upward SAW can appear in the downward SAW.

The motivation behind using such an energy-biased scheme is that it may allow for large changes of configuration and yet select final states that are "typical" of the system's target distribution. One can imagine, say, to uniformly perturb a large number of bits, but this is likely to yield states of high energy, and an MCMC algorithm will be extremely unlikely to accept the move at low temperatures. We will have more to say about the choice of the biasing parameter $\gamma$ as well as the related issue of the SAW lengths in Section 4.

Note that by multiplying the SAW flipping probabilities, we can straightforwardly obtain the probability of moving from state $\mathbf{x}_0$ to $\mathbf{x}_1$ *along the two SAWs* $(\boldsymbol{\sigma}, \boldsymbol{\rho})$, which we call $f(\mathbf{x}_1, \boldsymbol{\sigma}, \boldsymbol{\rho}|\mathbf{x}_0)$:

$$f(\mathbf{x}_1, \boldsymbol{\sigma}, \boldsymbol{\rho}|\mathbf{x}_0) \triangleq \delta_{\mathbf{x}_1}[F(\mathbf{x}_0, \boldsymbol{\sigma}, \boldsymbol{\rho})] \prod_{i=1}^{M} f_{up}(\sigma_i|\mathbf{u}_{i-1}) \times \prod_{i=1}^{M} f_{down}(\rho_i|\mathbf{d}_{i-1}) \quad (3)$$

The delta function simply enforces the fact that the final *state* $\mathbf{x}_1$ must result from the sequence of flips in $(\boldsymbol{\sigma}, \boldsymbol{\rho})$ from $\mathbf{x}_0$. The set of $\{\boldsymbol{\sigma}, \boldsymbol{\rho}\}$ such that $f(\mathbf{x}_1, \boldsymbol{\sigma}, \boldsymbol{\rho}|\mathbf{x}_0) > 0$ are termed the *allowable SAWs* between $\mathbf{x}_0$ and $\mathbf{x}_1$.

Ideally, to implement a Metropolis-Hastings (MH) algorithm using the SAW proposal, we would like to evaluate the *marginal* probability of proposing $\mathbf{x}_1$ from $\mathbf{x}_0$, which we call $f(\mathbf{x}_1|\mathbf{x}_0)$, so that the move would be accepted with the usual MH ratio:

$$\alpha_{mar}(\mathbf{x}_0, \mathbf{x}_1) \triangleq \min\left(1, \frac{\pi_n(\mathbf{x}_1)f(\mathbf{x}_0|\mathbf{x}_1)}{\pi_n(\mathbf{x}_0)f(\mathbf{x}_1|\mathbf{x}_0)}\right)$$

Unfortunately, for all but small values of the walk lengths $k$, marginalization of the proposal is intractable due to the potentially massive number of allowable SAWs between the two states.

To assist in illustrating our solution to this, we recall that a sufficient condition for a Markov transition kernel $K$ to have target $\pi_n$ as its stationary distribution is *detailed balance:*

$$\pi_n(\mathbf{x}_0)K(\mathbf{x}_1|\mathbf{x}_0) = \pi_n(\mathbf{x}_1)K(\mathbf{x}_0|\mathbf{x}_1) \quad (4)$$

One special case is obtained if we use the marginalized proposal $f(\mathbf{x}_1|\mathbf{x}_0)$ followed by the MH accept rule,

$$K_{mar}(\mathbf{x}_1|\mathbf{x}_0) \triangleq f(\mathbf{x}_1|\mathbf{x}_0)\alpha_{mar}(\mathbf{x}_0, \mathbf{x}_1) \quad (5)$$

As we cannot compute $f(\mathbf{x}_1|\mathbf{x}_0)$, we shall use a kernel $K(\mathbf{x}_1, \boldsymbol{\sigma}, \boldsymbol{\rho}|\mathbf{x}_0)$ defined on the *joint space* of SAWs and states, and show that with some care, detailed balance (4) can still *hold marginally.* It will be clear, though that this does *not* mean that the resultant marginal kernel $K(\mathbf{x}_1|\mathbf{x}_0)$ is the same as that in (5) obtained using MH acceptance on the marginal *proposal* .

Define the *sequence reversal operator* $R(\boldsymbol{\sigma})$ to simply return a sequence consisting of the elements of $\boldsymbol{\sigma}$ in reverse order; for example $R([2,3,1,4]) = [4,1,3,2]$. One can straightforwardly observe that each allowable up/down SAW pair $(\boldsymbol{\sigma}, \boldsymbol{\rho})$ from $\mathbf{x}_0$ to $\mathbf{x}_1$ can be uniquely mapped to the two allowable SAWs $(R(\boldsymbol{\rho}), R(\boldsymbol{\sigma}))$ from $\mathbf{x}_1$ to $\mathbf{x}_0$. For example in Figure 1, the SAWs $(R(\boldsymbol{\rho}) = [2,7,5], R(\boldsymbol{\sigma}) = [5,2,4])$ can be seen to be allowable from $\mathbf{x}_1$ to $\mathbf{x}_0$. Next, we have the following somewhat more involved concept:

**Definition 1.** *Consider a Markov kernel* $K(\mathbf{x}_1, \boldsymbol{\sigma}, \boldsymbol{\rho}|\mathbf{x}_0)$ *whose support set coincides with that of (3). We say that* pathwise detailed balance *holds if* $\pi_n(\mathbf{x}_0)K(\mathbf{x}_1, \boldsymbol{\sigma}, \boldsymbol{\rho}|\mathbf{x}_0) = \pi_n(\mathbf{x}_1)K(\mathbf{x}_0, R(\boldsymbol{\rho}), R(\boldsymbol{\sigma})|\mathbf{x}_1)$, *for all* $\boldsymbol{\sigma}, \boldsymbol{\rho}, \mathbf{x}_0, \mathbf{x}_1$.

It turns out that pathwise detailed balance is a *stronger condition* than marginal detailed balance. In other words,

**Proposition 1.** *If the property in Definition 1 holds for a transition kernel $K$ of the type described there, then* $\pi_n(\mathbf{x}_0)K(\mathbf{x}_1|\mathbf{x}_0) = \pi_n(\mathbf{x}_1)K(\mathbf{x}_0|\mathbf{x}_1)$

*Proof.* Suppose, for given $\mathbf{x}_0, \mathbf{x}_1$, we summed both sides of the equation enforcing pathwise detailed balance over all allowable SAWs $\{\boldsymbol{\sigma}', \boldsymbol{\rho}'\}$ from $\mathbf{x}_0$ to $\mathbf{x}_1$, i.e.

$$\sum_{\boldsymbol{\sigma}'\boldsymbol{\rho}'} \pi_n(\mathbf{x}_0)K(\mathbf{x}_1, \boldsymbol{\sigma}', \boldsymbol{\rho}'|\mathbf{x}_0) = \sum_{\boldsymbol{\sigma}'\boldsymbol{\rho}'} \pi_n(\mathbf{x}_1)K(\mathbf{x}_0, R(\boldsymbol{\rho}'), R(\boldsymbol{\sigma}')|\mathbf{x}_1)$$

The left-hand summation marginalizes the kernel over allowable SAWs and hence results in $\pi_n(\mathbf{x}_0)K(\mathbf{x}_1|\mathbf{x}_0)$. The observation above that each allowable SAW pair from $\mathbf{x}_0$ to $\mathbf{x}_1$ can be reversed to yield an allowable pair from $\mathbf{x}_1$ to $\mathbf{x}_0$ implies that the right-hand side is simply a re-ordered summation over all allowable SAWs from $\mathbf{x}_1$ to $\mathbf{x}_0$, and can thus be written as $\pi_n(\mathbf{x}_1)K(\mathbf{x}_0|\mathbf{x}_1)$. □

We are now ready to state the final form of the algorithm, which can be seen to instantiate a Markov chain satisfying pathwise detailed balance. After proposing $(\mathbf{x}_1, \boldsymbol{\sigma}, \boldsymbol{\rho})$ using the up/down joint process, we accept the move with the ratio:

$$\alpha(\mathbf{x}_0, \mathbf{x}_1, \boldsymbol{\sigma}, \boldsymbol{\rho}) \triangleq \min\left(1, \frac{\pi_n(\mathbf{x}_1)f(\mathbf{x}_0, R(\boldsymbol{\rho}), R(\boldsymbol{\sigma})|\mathbf{x}_1)}{\pi_n(\mathbf{x}_0)f(\mathbf{x}_1, \boldsymbol{\sigma}, \boldsymbol{\rho}|\mathbf{x}_0)}\right) \quad (6)$$

The computational complexity of evaluating this accept ratio is of the same order as that required to sample the proposed SAWs/state; the only additional operations required are those needed to evaluate the reverse proposal appearing in the numerator, which are completely analogous to those involved in calculating the forward proposal.

Before proceeding to the experimental validation, let us take a closer look at the marginal transition kernel $K(\mathbf{x}_1|\mathbf{x}_0)$. We can factor the joint *proposal* into:

$$f(\mathbf{x}_1, \boldsymbol{\sigma}, \boldsymbol{\rho}|\mathbf{x}_0) = f(\mathbf{x}_1|\mathbf{x}_0)f(\boldsymbol{\sigma}, \boldsymbol{\rho}|\mathbf{x}_0, \mathbf{x}_1)$$

Of course, if we are assuming that $f(\mathbf{x}_1|\mathbf{x}_0)$ is intractable to evaluate, then the conditional $f(\boldsymbol{\sigma}, \boldsymbol{\rho}|\mathbf{x}_0, \mathbf{x}_1)$ must be so as well, but it is useful to consider. If we now summed both sides of the *joint* probability of moving from $\mathbf{x}_0$ to $\mathbf{x}_1$ over allowable paths, we would observe:

$$\sum_{\boldsymbol{\sigma}'\boldsymbol{\rho}'} \pi_n(\mathbf{x}_0) K(\mathbf{x}_1, \boldsymbol{\sigma}', \boldsymbol{\rho}'|\mathbf{x}_0) =$$
$$\pi_n(\mathbf{x}_0) f(\mathbf{x}_1|\mathbf{x}_0) \sum_{\boldsymbol{\sigma}'\boldsymbol{\rho}'} f(\boldsymbol{\sigma}', \boldsymbol{\rho}'|\mathbf{x}_0, \mathbf{x}_1) \alpha(\mathbf{x}_0, \mathbf{x}_1, \boldsymbol{\sigma}', \boldsymbol{\rho}')$$

The summation on the right-hand side is thus the *conditional expectation* of the accept rate given that we are attempting to move from $\mathbf{x}_0$ to $\mathbf{x}_1$; we call it

$$\alpha(\mathbf{x}_0, \mathbf{x}_1) \triangleq \sum_{\boldsymbol{\sigma}'\boldsymbol{\rho}'} f(\boldsymbol{\sigma}', \boldsymbol{\rho}'|\mathbf{x}_0, \mathbf{x}_1) \alpha(\mathbf{x}_0, \mathbf{x}_1, \boldsymbol{\sigma}', \boldsymbol{\rho}')$$

and it defines an *effective acceptance rate* between $\mathbf{x}_0$ and $\mathbf{x}_1$ under the sampling regime described since $K(\mathbf{x}_1|\mathbf{x}_0) = f(\mathbf{x}_1|\mathbf{x}_0)\alpha(\mathbf{x}_0, \mathbf{x}_1)$. Clearly, $\alpha(\mathbf{x}_0, \mathbf{x}_1) \neq \alpha_{mar}(\mathbf{x}_0, \mathbf{x}_1)$, *i.e.* the marginal accept rate for the joint proposal is not the same as the one we get when using the marginalized proposal.

Before proceeding to the experiments, we briefly touch on some implementation considerations; a detailed discussion will appear in a longer report. The bulk of the computational time of the IM algorithm is spent in generating states with the SAW proposal. At each step of the process, a component from a discrete probability vector, corresponding to the variable to flip, must be sampled. Naively, the time needed to do so scales linearly with the length $l$ of the vector. In graphical models of sparse connectivity, however, it is possible to achieve a dramatic computational speedup by storing the vector in a *binary heap.* Sampling from a heap is of $O(\log l)$, but for *sparsely connected* models, updating the heap in response to a flip, which entails replacing the energy changes that would result if the flipped variable's *neighboring* variables were themselves to flip, is also of logarithmic complexity. In contrast, for a densely connected model, the heap update would be of $O(M \log l)$, while recomputing the vector in the naive method is $O(M)$. The simple method is thus cheaper for dense models.

## 4 EXPERIMENTS

Our experimental validation will consist of comparing the popular Metropolis algorithm defined on the restricted sets discussed in Section 2 to the IM sampler. Three types of binary-valued systems, all belonging to the general class of undirected graphical model called the *Ising model*, were used. The energy of a binary state $\mathbf{s}$, where $s_i \in \{-1, 1\}$ is given by:

$$E(\mathbf{s}) = -\sum_{(i,j)} J_{ij} s_i s_j - \sum_i h_i s_i$$

(One can trivially map $x_i \in \{0, 1\}$ to $s_i \in \{-1, 1\}$ and vice-versa.) The interaction weights $J_{ij}$ between variables $i$ and $j$ are zero if they are topologically disconnected; positive (also called "ferromagnetic") if they tend to have the same value; and negative ("anti-ferromagnetic") if they tend to have opposite values. The presence of interactions of mixed sign, as routinely occurs in Boltzmann machines, restricted or otherwise, can significantly complicate Monte Carlo simulation due to the proliferation of local minima in the energy landscape. Interaction weights of different sign produce unsatisfiable constraints and cause the system to become "frustrated".

The first set of experiments will consider the behavior of the two algorithms on a ferromagnetic Ising model on a planar, regular grid of size $60 \times 60$, where all $J_{ij}$ were 1 and all $h_i$ were 0. The target set $\mathcal{S}_n$ was the collection of all states with exactly half of the $\{s_i\}$ having values of 1 and the remaining ones having values of $-1$. $\beta$ was set to $1/2.27$, corresponding to the so-called *critical temperature*, where many interesting phenomena arise (Newman & Barkema 1999) but where simulation also becomes quite difficult. It is worth mentioning that the set constraint makes it not at all obvious how to apply methods known do do well in the unconstrained ferromagnet case, such as variational inference and Swendsen-Wang.

The second batch of experiments will compare the algorithms on an Ising model where the variables are topologically structured as a $9\times9\times9$ three-dimensional cube, $J_{ij}$ are uniformly sampled from the set $\{-1, 1\}$, and the $h_i$ are zero. $\beta$ was set to 1.0, corresponding to a lower temperature than the value of 0.9, at which it is known (Marinari, Parisi & Ruiz-Lorenzo 1997) that, roughly speaking, regions of the state space become very difficult to visit from one another via traditional Monte Carlo simulation. $\mathcal{S}_n$ was specified to be the set of states with 364 of the 729 $s_i = 1$.

While the three-dimensional-cube spin-glass is a much harder problem than the ferromagnet, it represents a worst case scenario. One would hope that problems arising in practice will have structure in the potentials that would ease the problem of inference. For this reason, the final experimental set consisted of runs on a restricted Boltzmann machine (Smolensky 1986) with a constraint on the number of active units. RBMs are bipartite undirected probabilistic graphical models. The variables on one side are often referred to as "visible units", while the others are called "hidden units". Each visible unit is connected to all hidden units. However there are no connections among the hidden units and among the visible units. Therefore, given the visible units, the hidden units are conditionally independent and vice-versa. Our model consisted of 784 and 500 visible and hidden units respectively. We chose weights corresponding to local Gabor filters to purposely capture regularities corresponding to the natural statistics of many perceptual inputs, such as natural images (Hyvarinen, Hurri & Hoyer 2009). The parameter $\beta$ was set to one, and $\mathcal{S}_n$ was the set of states with 1/3 of the $\{s_i\}$ having value 1. The total number of variables and edges in the graph were thus 1484 and 392000 respectively. It is worth mentioning that the constraints prohibit us from using the standard blocked Gibbs samplers that exploit conditional independence for inference in standard RBMs.

The experimental protocol was roughly the same for all models: For 10 independent trials, run the two samplers for a certain number of iterations, storing the sequence of energies visited. Using each trial's energy sequence, compute the *autocorrelation function* (ACF.) Comparison of the two algorithms consisted of analyzing the energy ACF averaged over the trials. Without going into detail, a more rapidly decreasing ACF is indicative of a faster-mixing Markov chain; see for example (Robert & Casella 2004).

Before moving on to the results, it is crucial to discuss the issue of fairness in comparing the constant-set Metropolis algorithm to the IM sampler. One concern is that the SAW proposal takes considerably more computational time to propose a candidate state than the Metropolis algorithm's method of uniformly flipping and "unflipping" a variable. In all our experiments, we systematically allowed Metropolis to have more iterations so that it consumed the same computational time as IM at a given set of parameters. To compute the energy ACF generated by Metropolis, however, we subsampled the energy sequence: For example, if a fair comparison necessitated that Metropolis be given ten times more iterations than those given to IM, every tenth sample from the Metropolis' output sequence was taken when computing the ACF. Thus, as a just comparison requires, we calculated the ACF *as a function of computational time lag* instead of as a function of Monte Carlo iteration lag, the latter which may unfairly inflate the correlation of the computationally cheaper Metropolis algorithm.

For the first two models, which were of sparse connectivity, the SAW proposals were implemented using the efficient binary heap procedure discussed in Section 3.

Figure 2 displays the results of the two samplers on the ferromagnetic Ising model at the critical temperature. For IM, the SAW length of the proposal was set to 90, and the up/down SAW biases $\gamma$ were set to the system's value of $\beta$. $10^5$ moves were attempted. To allow for computational parity, the Metropolis algorithm was given $5 \times 10^6$ move attempts. The progress of the energy generated by the two samplers is shown in Figure 2 (top); both samplers are exploring the same range of the energy space, but IM's energy trajectory is clearly less correlated than that of the Metropolis algorithm at a given point *in computational time*. The algorithms' ACFs shown below show a marked decrease in the mixing time when using IM.

We now present the results on the more challenging case of the spin-glass. In fact this set of experiments illustrates some of the potential issues in using the IM algorithm (or any other algorithm) in this challenging

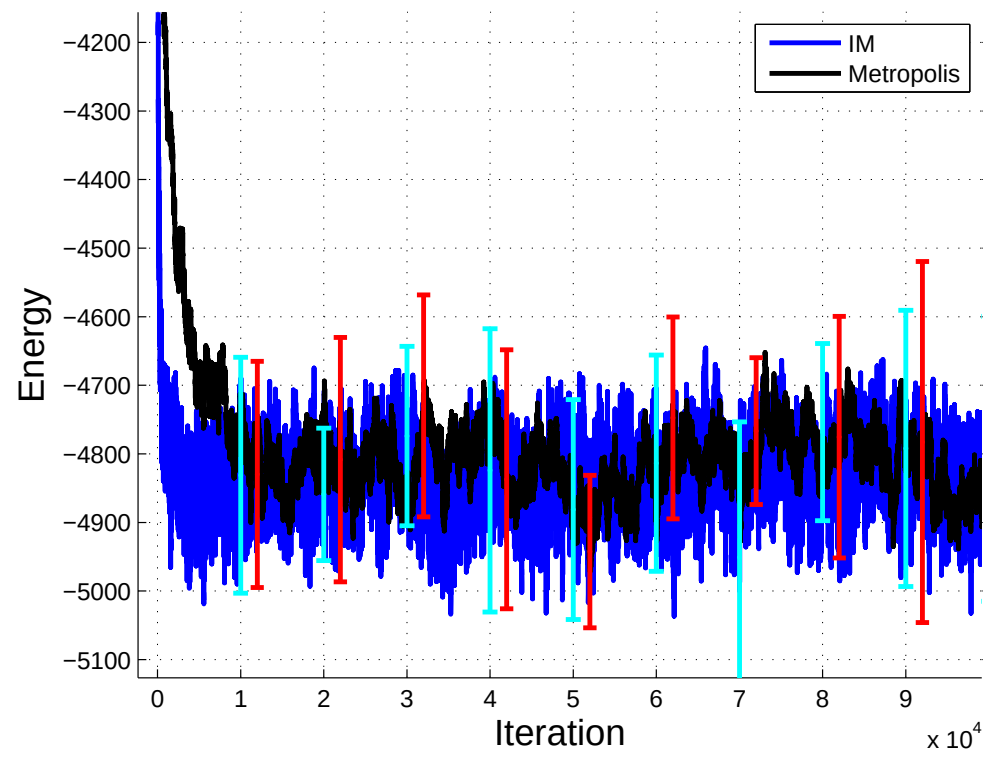

(a)

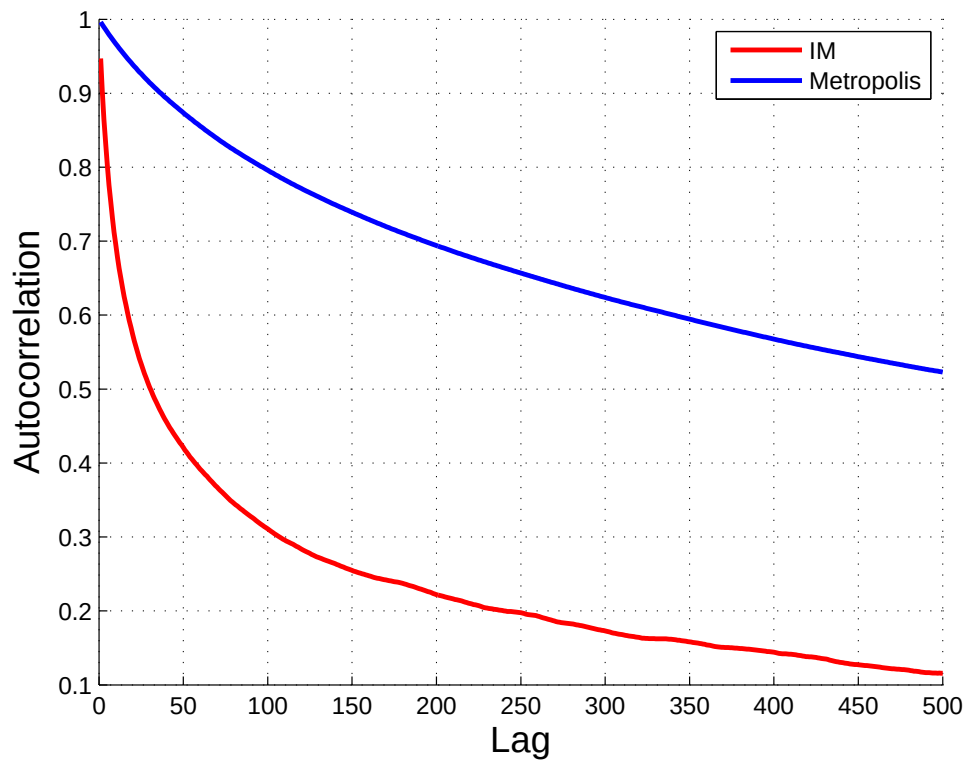

(b)

Figure 2: (Color online) Energy trajectory (a) and autocorrelation function (b) of the IM and Metropolis algorithms on the 2D $60 \times 60$ ferromagnetic Ising model as a function of *computational time*. Bars show the variance of the samples over 10 trials and are to aid in visualizing the spread of the samples generated at the corresponding points, not as "errors." See text for discussion of results and on how a fair comparison was made.

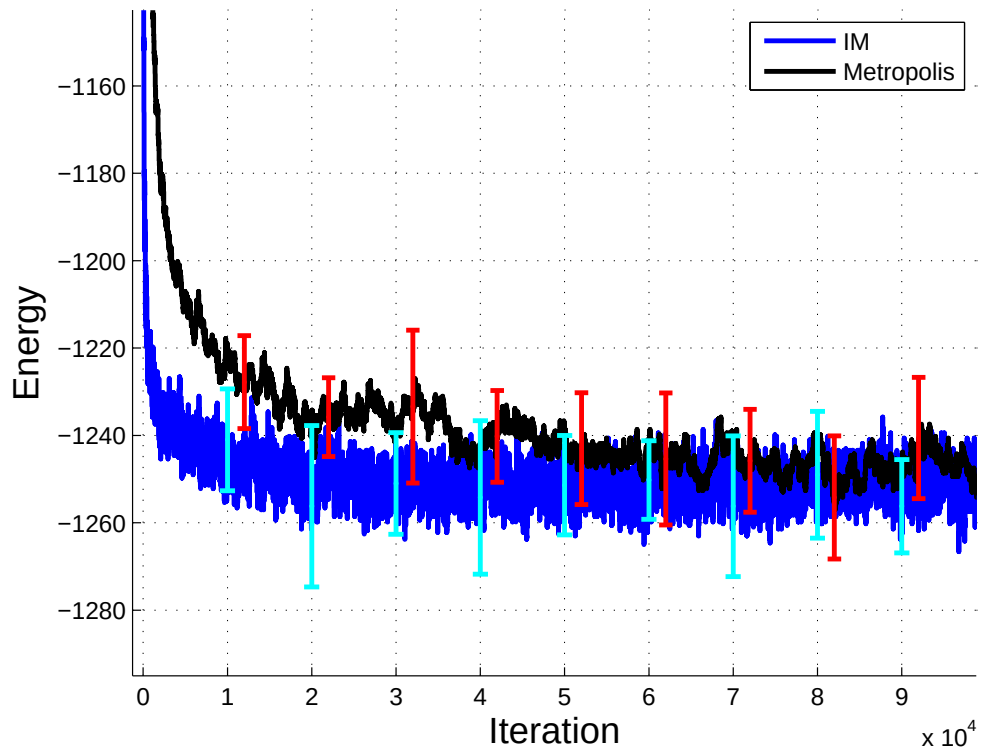

(a)

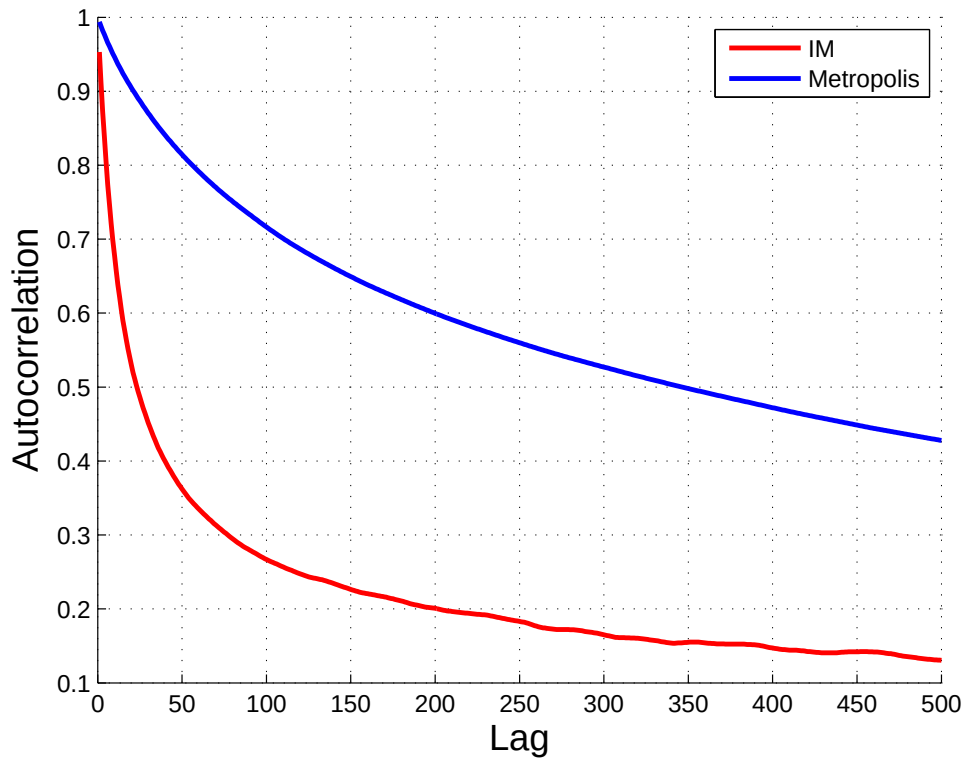

(b)

Figure 3: (Color online) Energy trajectory (a) and autocorrelation function (b) of the IM and Metropolis algorithms on the 3D $9 \times 9 \times 9$ Ising spin-glass as a function of *computational time*. Bars show the variance of the samples over 10 trials and are to aid in visualizing the spread of the samples generated at the corresponding points, not as "errors." The Metropolis algorithm is clearly not converging to the same statistical ensemble as that reached by IM.

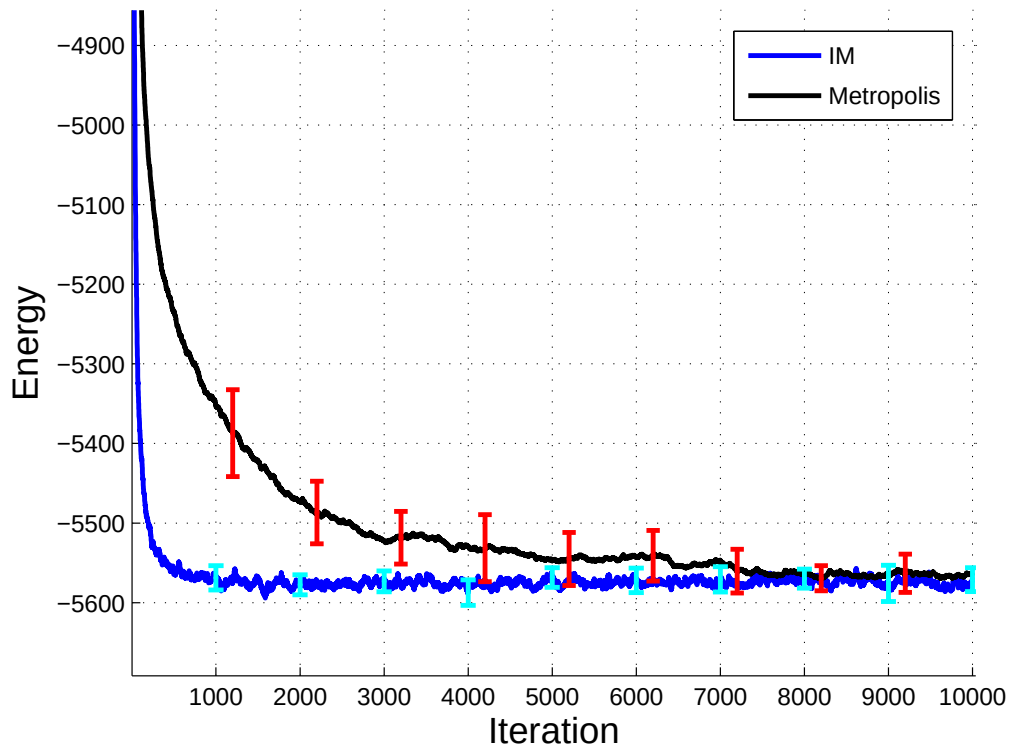

(a)

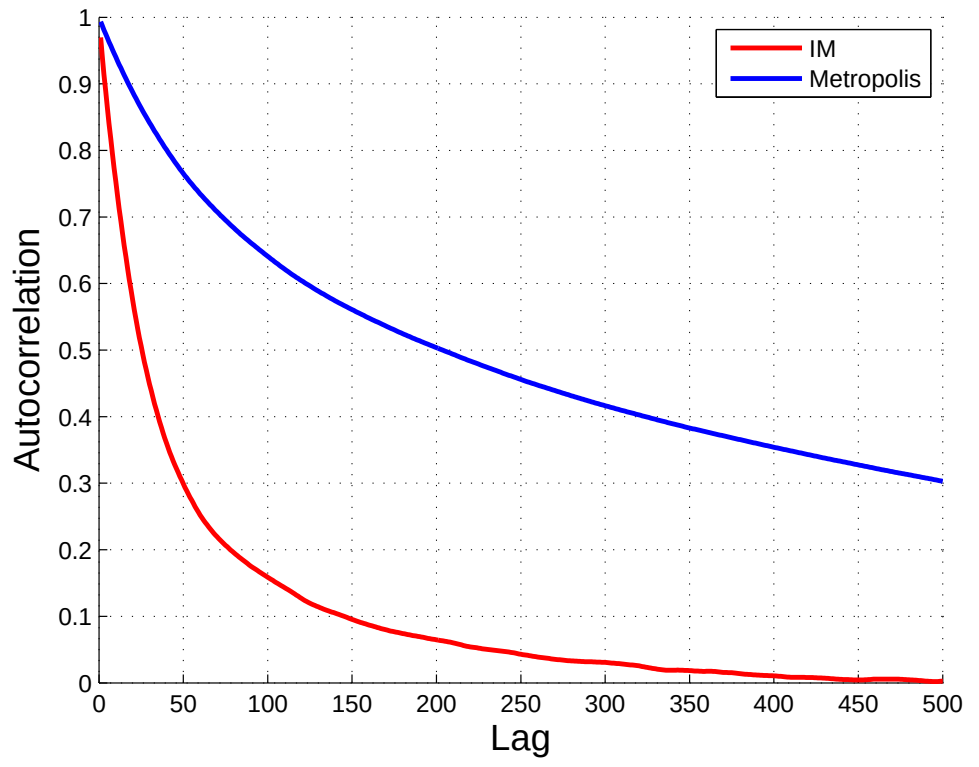

(b)

Figure 4: (Color online) Energy trajectory (a) and energy autocorrelation function (b) of the IM and Metropolis algorithms on the RBM Ising model as a function of *computational time.* Bars show the variance of the samples over 10 trials and are to aid in visualizing the spread of the samples generated at the corresponding points, not as "errors." See text for discussion of results and on how a fair comparison was made.

problem. Unlike the case of the ferromagnetic model, we found that using a single SAW length for the duration of the simulation was not ideal; we uniformly sampled a SAW length in the range $[1, 25]$ prior to each move attempt. Additionally, the SAW bias parameter $\gamma$ was set to $0.8\beta = 0.8$. This latter choice, along with the shorter SAW lengths relative to those used in the ferromagnetic model, are consequences of the extremely rugged energy landscape of these spin glass models and merit some closer consideration. We recall that in IM, the move is accepted with probability given by Equation 6, which is a function of the difference in energies of the initial and final states, but crucially in this case, also of the ratio of the reverse to forward path proposal probabilities. It turns out that for spin glasses, if the SAW length is too long, it becomes exceedingly *less* likely that state $\mathbf{s}_0$ is sampled from $\mathbf{s}_1$ along the reverse SAW to the one that resulted in $\mathbf{s}_1$ from $\mathbf{s}_0$. In other words, a SAW from $\mathbf{s}_1$ would rather go elsewhere than back to $\mathbf{s}_0$. Unless the energy of $\mathbf{s}_1$ is so low (relative to that of $\mathbf{s}_0$) so as to overwhelm this effect, moves will then be mostly rejected. Setting $\gamma$ to be too large has a similar effect for the same reason. Conversely, letting $\gamma$ to be too *low* will yield high rejection due to the high energy of the final state. The SAW lengths we used and setting $\gamma$ to be slightly below the one at which the state space becomes difficult to explore were a good trade-off that allowed the sampler to function satisfactorily.

A total of $10^5$ moves were attempted by IM, while $10^6$ were given to Metropolis. Figure 3 shows the results. We can see once again from both the plots of the energy and of the ACF that IM mixes considerably faster than the Metropolis algorithm. Indeed from the energy plot one can see that the Metropolis algorithm has not settled into the same statistical ensemble as that of IM by the end of the simulation, while at least in the ferromagnet, despite the high sample correlation, the Metropolis algorithm appears to have equilibrated. Thus in the ferromagnet, due to the high correlation Metropolis is giving fewer "effective" samples from the target than IM, but correct ones, while in the spin glass, Metropolis is not yet even sampling the same distribution as the one reached by IM, which is a much more serious problem.

Finally, Figure 4 shows the simulation results for the RBM. The SAW length for IM was uniformly sampled prior to each move attempt in the range $[1, 20]$, $\gamma = 0.8$. $10^4$ and $10^5$ moves were attempted by the IM and Metropolis algorithms respectively. We can see again that the performance of IM is markedly superior to that of the Metropolis algorithm; as in the case of the spin glass, the ensemble sampled by Metropolis has not converged to the one that IM reached quite early in the simulation. The parameters we have chosen cause an especially rapid drop-off in the ACF of IM. Please note the vertical scale in the autocorrelation plots. In the RBM, it decreases quickly to 0 and not 0.1 as in the previous models. This seems to indicate that potentials corresponding to regularities in perceptual signals (*e.g.* natural image statistics) can be easier to treat during the inference process.

# 5 CONCLUSION

We presented a novel MCMC method for sampling from binary distributions with particular types of constraints. The experiments demonstrated, in three Ising models of varying difficulty, that the new sampler outperforms the popular algorithm in the literature by a significant margin. Energy and autocorrelation plots, showed that the new sampler converges faster to the stationary regime and that it produces more independent samples. Although we focused on Ising models for demonstration purposes, the algorithm applies to arbitrary binary distributions. Testing the algorithm in these other distributions is a good avenue for future work. We tuned the SAW length and the proposal parameter $\gamma$ to obtain good acceptance rates. Although this is fairly easy to do, it would also be a good idea to do this automatically using standard adaptive MCMC tools.